\documentclass[twocolumn,preprintnumbers,amsmath,amssymb,prl]{revtex4}
\usepackage{graphicx}
\usepackage{dcolumn}
\usepackage{bm}
\usepackage{units}%
\usepackage{stmaryrd}%
\usepackage[breaklinks=true,pdfborder={0 0 0}]{hyperref}
\usepackage[usenames]{color}
\usepackage{tabularx}

\usepackage{color}
\usepackage{tikz}
\usepackage{multirow}
\usepackage{bbold}
\usepackage{accents}
\hypersetup{colorlinks=true,citecolor=red}
\DeclareMathAlphabet\mathbfcal{OMS}{cmsy}{b}{n}
\begin{document}
\newlength{\LL} \LL 1\linewidth
\title{Four-Dimensional Scaling of Dipole Polarizability in Quantum Systems}
\author{P\'eter~Szab\'o}
\email[E-mail address: ]{peter88szabo@gmail.com}
\affiliation{Department of Physics and Materials Science, University of Luxembourg, L-1511 Luxembourg City, Luxembourg}
\author{Szabolcs~G\'oger}
\affiliation{Department of Physics and Materials Science, University of Luxembourg, L-1511 Luxembourg City, Luxembourg}
\author{Jorge Charry}
\affiliation{Department of Physics and Materials Science, University of Luxembourg, L-1511 Luxembourg City, Luxembourg}
\author{Mohammad~Reza~Karimpour}
\affiliation{Department of Physics and Materials Science, University of Luxembourg, L-1511 Luxembourg City, Luxembourg}
\author{Dmitry~V.~Fedorov}
\email[E-mail address: ]{dmitry.fedorov@uni.lu}
\affiliation{Department of Physics and Materials Science, University of Luxembourg, L-1511 Luxembourg City, Luxembourg}
\author{Alexandre~Tkatchenko}
\email[E-mail address: ]{alexandre.tkatchenko@uni.lu}
\affiliation{Department of Physics and Materials Science, University of Luxembourg, L-1511 Luxembourg City, Luxembourg}


\begin{abstract}
Polarizability is a key response property of physical and chemical systems, which has an impact on intermolecular interactions,
spectroscopic observables, and vacuum polarization. The calculation of polarizability for quantum systems involves an infinite sum
over all excited (bound and continuum) states, concealing the physical interpretation of polarization mechanisms and complicating
the derivation of efficient response models. Approximate expressions for the dipole polarizability, $\alpha$, rely on different
scaling laws $\alpha \propto$ $R^3$, $R^4$, or $R^7$, for various definitions of the system radius $R$. Here, we consider a range
of single-particle quantum systems of varying spatial dimensionality and having qualitatively different spectra, demonstrating
that their polarizability follows a universal four-dimensional scaling law $\alpha = C (4 \mu q^2/\hbar^2)L^4$, where $\mu$ and
$q$ are the (effective) particle mass and charge, $C$ is a dimensionless excitation-energy ratio, and the characteristic length
$L$ is defined via the $\mathcal{L}^2$-norm of the position operator. This unified formula is also applicable to many-particle
systems, as shown by accurately predicting the dipole polarizability of 36 atoms, 1641 small organic molecules, and Bloch
 electrons in periodic systems.
\end{abstract}

\maketitle
The dipole polarizability determines the strength of the response of a system of charged particles to applied electric fields as well as dispersion/polarization interactions between
atoms or molecules~\cite{Stone-book,Atkins&Friedman-book,Hermann2017},
playing an important role in the interpretation of 
experiments~\cite{Wang2006,Dakovski2007,Seufert2001,Empedocles1997,Kulakci2008}. Efficient models for polarizability are useful
to predictively describe various phenomena in physics, chemistry,
and biology. Moreover, a detailed understanding of quantum-mechanical (QM) polarization mechanisms could help in developing a microscopic picture
of intrinsic vacuum response properties~\cite{Milonni-book,Leuchs2013,Urban2013}. 
In general, the dipole polarizability is a second-rank
tensor which determines the dipole moment induced by an applied electric field: $\mathbf{d} = \tensor{\alpha} \mathbfcal{E}$.
For anisotropic systems, the polarizability tensor can be diagonalized using the principal axes, whereas in the case of isotropic systems it effectively reduces to a scalar: $\alpha_{ii} = \alpha = \frac{1}{3}\, {\rm Tr}\, \tensor{\alpha}$. For a QM system in its ground state, the dipole
polarizability can be  evaluated via the Rayleigh-Schr\"odinger perturbation
theory~\cite{Atkins&Friedman-book}
\begin{align}
\tensor{\alpha} = 2\textstyle\sum\limits^{\infty}_{n\neq 0}
{\langle \varPsi_{0}|\hat{\mathbf{d}}|\varPsi_{n}\rangle \otimes \langle \varPsi_{n}|\hat{\mathbf{d}}|\varPsi_{0}\rangle }/{(E_{n}-E_{0})}\ ,
\label{eq.:alpha1_perturb_def}
\end{align}
where $\otimes$ indicates
the dyadic vector product and the sum goes over all excited states.
This formula describes transient or fluctuating electric dipoles 
as the matrix elements of the dipole operator 
$\hat{\mathbf{d}} = \sum_j \hat{\mathbf{d}}_j = \sum_j q_j \hat{\mathbf{r}}_j$, where $q_j$ and $\hat{\mathbf{r}}_j$
are the charge and position operator of the $j$th particle, respectively.
For an accurate calculation of $\alpha$, all bound and continuum states must be taken into account. 
Thus, Eq.~(\ref{eq.:alpha1_perturb_def}), while being exact, is difficult to evaluate in practice. Therefore, various
approximations~\cite{Unsold1927,Vinti1932,Kirkwood1932,Buckingham1979,Montgomery2013} have been developed for a more efficient evaluation of Eq.~(\ref{eq.:alpha1_perturb_def}). Besides their computational advantage, approximate models often provide a deeper insight into the polarizability and its relation to other physical observables.

According to Eq.~(\ref{eq.:alpha1_perturb_def}), the polarizability should be related to a certain characteristic length for a given QM system. This has led to a proposition of a number of scaling laws with respect to different effective system sizes:
\begin{align}
\alpha \propto R_{\rm cl}^3\ ,\ \ \alpha \propto R_{\rm conf}^4\ ,\ \ \alpha \propto R_{\rm vdW}^7\ .
\label{eq.:three_SL}
\end{align}
The first relation stems from the classical formula, $\alpha = (4\pi\epsilon_0) R_{\rm cl}^3$, where $\epsilon_0$ is the vacuum permittivity and $R_{\rm cl}$ is the radius of a conducting spherical shell~\cite{Hirschfelder-book} or a hard sphere with uniform electron density and a positive point charge at its center~\cite{Griffiths-book}. This formula delivers the most commonly accepted scaling law, which is used in practice to describe the polarizability of atoms in molecules and materials~\cite{Johnson2005, Mayer2007, Tkatchenko2009, DeKock2012, DelloStritto2019}. 
The second relation in Eq.~\eqref{eq.:three_SL} holds for confined quantum systems of length $R_{\rm conf}$, as was derived by Fowler~\cite{Fowler1984}. This relation was observed
for semiconductor nanocrystals by using terahertz time-domain spectroscopy~\cite{Wang2006,Dakovski2007}.
The third scaling law, $\alpha \propto R_{\rm vdW}^7\,$, connecting the atomic polarizability and van der Waals (vdW) radius, was found~\cite{Fedorov2018,Tkatchenko2020} by studying the balance between exchange and correlation forces for two interacting quantum Drude oscillators (QDO)~\cite{Wang2001,Sommerfeld2005,Jones2013}. The approach of Ref.~\cite{Fedorov2018} has been subsequently employed to improve effective models for vdW interactions~\cite{Silvestrelli2019,Rudden2019}.
All the three distinct scaling laws can be represented as
\begin{align}
\alpha = (4\pi\epsilon_0) R_p^3\, (R_p/R_{p}^{r})^p\ ,
\label{eq.:three_in_one}
\end{align}
where $(R_p/R_{p}^{r})^p$ is a correction to the classical formula. The renormalization length $R_{p}^{r}$ depends on the choice of the effective system size $R_p \in \{R_{\rm cl}, R_{\rm conf}, R_{\rm vdW}\}$, that corresponds to $p = \{0, 1, 4\}$.
Whereas $R_{1}^{r}$ depends on the system parameters~\cite{Fowler1984}, $R_{4}^{r}$ was found~\cite{Fedorov2018,Tkatchenko2020} to be the same for all atoms. However, the vdW radius is an interacting radius rather than an effective system size and its accurate evaluation independent from the polarizability is difficult~\cite{Fedorov2018}. Therefore, it is desirable to establish a general relation of $\alpha$ to a concrete effective size of any given QM system, such as the scaling law for confined systems with a defined confinement radius~\cite{Fowler1984}. Since Eq.~\eqref{eq.:three_in_one} gives the right units of $\alpha$ for any value of $p$, the form of such a general relation is not obvious \emph{a priori}. 

In this Letter, we show that for distinct QM systems
the principal-axis components of the polarizability tensor in Eq.~(\ref{eq.:alpha1_perturb_def}) are given by a unified expression
\begin{align}
\alpha_{ii} = C_i ({4 \mu q^2}/{\hbar^2}) L_i^{4}\ ,
\label{eq.:alpha_L4}
\end{align}
where the constant $C_i$ depends on properties of the quantum particle with mass $\mu$ and charge $q$.
The characteristic length $L_i$ measures the spatial spread of the ground-state wave function $\Psi_0$ with respect to its center
of mass $\mathbf{R} = (R_1, R_2, ..., R_N) = \langle \Psi_0 |\hat{\mathbf{r}}| \Psi_0 \rangle$, which corresponds to the nuclear position in case of atoms.
The Euclidean $\mathcal{L}^2$-norm of the position
vector, $(\mathbf{r}-\mathbf{R})$, is defined for a QM system described by its ground-state wavefunction as
\begin{align}
L_i = \sqrt{\textstyle\int {\, (r_i - R_i)^2 \, |\Psi_0 (\mathbf{r})|^2} \, \rm d\mathbf{r}^\emph{N}} \ ,
\label{eq.:L_definition}
\end{align}
where $N$ is the system spatial dimensionality (D).
Equation~(\ref{eq.:alpha_L4}), connecting $\alpha_{ii}$ with the characteristic length $L_i$ along the $i$th principal axis, makes our
approach applicable to QM systems of any dimensionality.
Moreover, for atom-like systems with a well-defined positively charged center of mass, the dimensionless constant $C_i$ turns out to be close to unity. Equation~\eqref{eq.:alpha_L4} scales as the relation obtained by Fowler~\cite{Fowler1984} for confined systems via either an exact derivation of the polarizability or its Uns\"old~\cite{Unsold1927} and Kirkwood~\cite{Kirkwood1932} estimates. However, the size of such confined systems was imposed as a classical parameter which cannot be defined for QM systems in free space. As we show below, with our choice of the characteristic length -- a QM generalization of the conventional Euclidean $\mathcal{L}^2$-norm -- one can properly describe the polarizability of any atom-like QM system.

To demonstrate the general validity of Eq.~(\ref{eq.:alpha_L4}), we start with the approach of Vinti~\cite{Vinti1932}, which bridges the Uns\"old and Kirkwood approximations, and employ the Integral Mean Value Theorem (IMVT)~\cite{IMVT1} for the polarizability
\begin{align}
\alpha_{ii}=({2q^{2}}/{\Delta E_i})
\textstyle\sum\limits_{n>0}^{\infty}\left\langle
\varPsi_{0}\left|\hat{r}_i\right|\varPsi_{n}\right\rangle \left\langle \varPsi_{n}\left|\hat{r}_i\right|\varPsi_{0}\right\rangle\ ,
\label{eq.:alpha1_Unsold}
\end{align}
by introducing $\Delta E_i$ as an effective excitation energy. Based on the IMVT, $\Delta E_i$ can be chosen to give the exact polarizability,  
which differs from the Uns\"old approximation~\cite{Unsold1927} with 
$\Delta E = E_1 - E_0$ providing an upper bound estimate for the polarizability $\left(\alpha \leq \alpha^U\right)$, as was proven variationally by Fowler \cite{Fowler1984}.
The IMVT allows us to connect the polarizability to the variance of the position operator,
$\left(\Delta r_i\right)^{2}=\left\langle r_i^{2}\right\rangle -\left\langle r_i\right\rangle ^{2}$, where
$\left\langle r_i^{2}\right\rangle =\left\langle \varPsi_{0}\left|\hat{r}_i^{2}\right|\varPsi_{0}\right\rangle$
and
$\left\langle r_i\right\rangle =\left\langle \varPsi_{0}\left|\hat{r}_i\right|\varPsi_{0}\right\rangle$.
Indeed, by using the closure relation~
$\sum_{n>0}^{\infty}\left|\varPsi_{n}\right\rangle \left\langle \varPsi_{n}\right|
= \mathbb{1}-\left|\varPsi_{0}\right\rangle \left\langle \varPsi_{0}\right|\,$, Eq.~(\ref{eq.:alpha1_Unsold}) reduces~to
\begin{align}
\alpha_{ii}=({2q^{2}}/{\Delta E_i})\left(\Delta r_i\right)^{2}\ .
\label{eq.:alpha1_deltaL4}
\end{align}
Employing now the IMVT~\cite{IMVT1} for the Thomas-Reiche-Kuhn (TRK) sum-rule~\cite{Wang1999}
\begin{align}
({2\mu}/\hbar^{2})\textstyle\sum\limits_{n>0}^{\infty}
\left(E_{n}-E_{0}\right)\left\langle \varPsi_{0}\left|\hat{r}_i\right|
\varPsi_{n}\right\rangle^{2}=1\ ,
\label{eq.:TRKSumrule}
\end{align}
we obtain another effective excitation energy as $(\Delta \widetilde{E}_i)^{-1}= ({2\mu}/{\hbar^{2}}) \left(\Delta r_i\right)^{2}$ which fulfills the exact TRK sum-rule. 
Generally, $\Delta \widetilde{E}_i$ is not equal to $\Delta E_i$ in Eq.~\eqref{eq.:alpha1_Unsold} but
there is a constant $C_i$ such that $\Delta \widetilde{E}_i = C_i \, \Delta E_i$. Inserting $(\Delta E_i)^{-1}= C_i\,({2\mu}/{\hbar^{2}}) \left(\Delta r_i\right)^{2}$ into Eq.~(\ref{eq.:alpha1_deltaL4}) yields 
\begin{align}
\alpha_{ii} = C_i\,({4 \mu q^{2}}/{\hbar^{2}}) \left(\Delta r_i\right)^{4}
= C_i\,({4 \mu q^{2}}/{\hbar^{2}})\, L_i^4\ ,
\label{eq.:alphaU_deltar4}
\end{align}
where we used the fact that $L_i^2$ defined via
Eq.~\eqref{eq.:L_definition} is identical to the variance, $(\Delta r_i)^{2}$. With $C_i = 1$ in Eq.~\eqref{eq.:alphaU_deltar4}, \emph{i.e.} $\Delta \widetilde{E} = \Delta E$, we reproduce Vinti's original derivation~\cite{Vinti1932} of the well-known Kirkwood formula~\cite{Kirkwood1932}, which yields a lower bound to the exact polarizability $\left(\alpha^K \leq \alpha\right)$~\cite{Fowler1984}.
The general formula, Eqs.~\eqref{eq.:alpha_L4} and \eqref{eq.:alphaU_deltar4}, is based on fundamental properties of QM systems caused by quantum fluctuations, which determine and relate $L_i$ in Eq.~(\ref{eq.:L_definition}) as the ground-state metric of the position operator to $\alpha_{ii}$ as determined by the transient electric dipoles in Eq.~(\ref{eq.:alpha1_perturb_def}).

\begin{figure*}
\includegraphics[width=18cm]{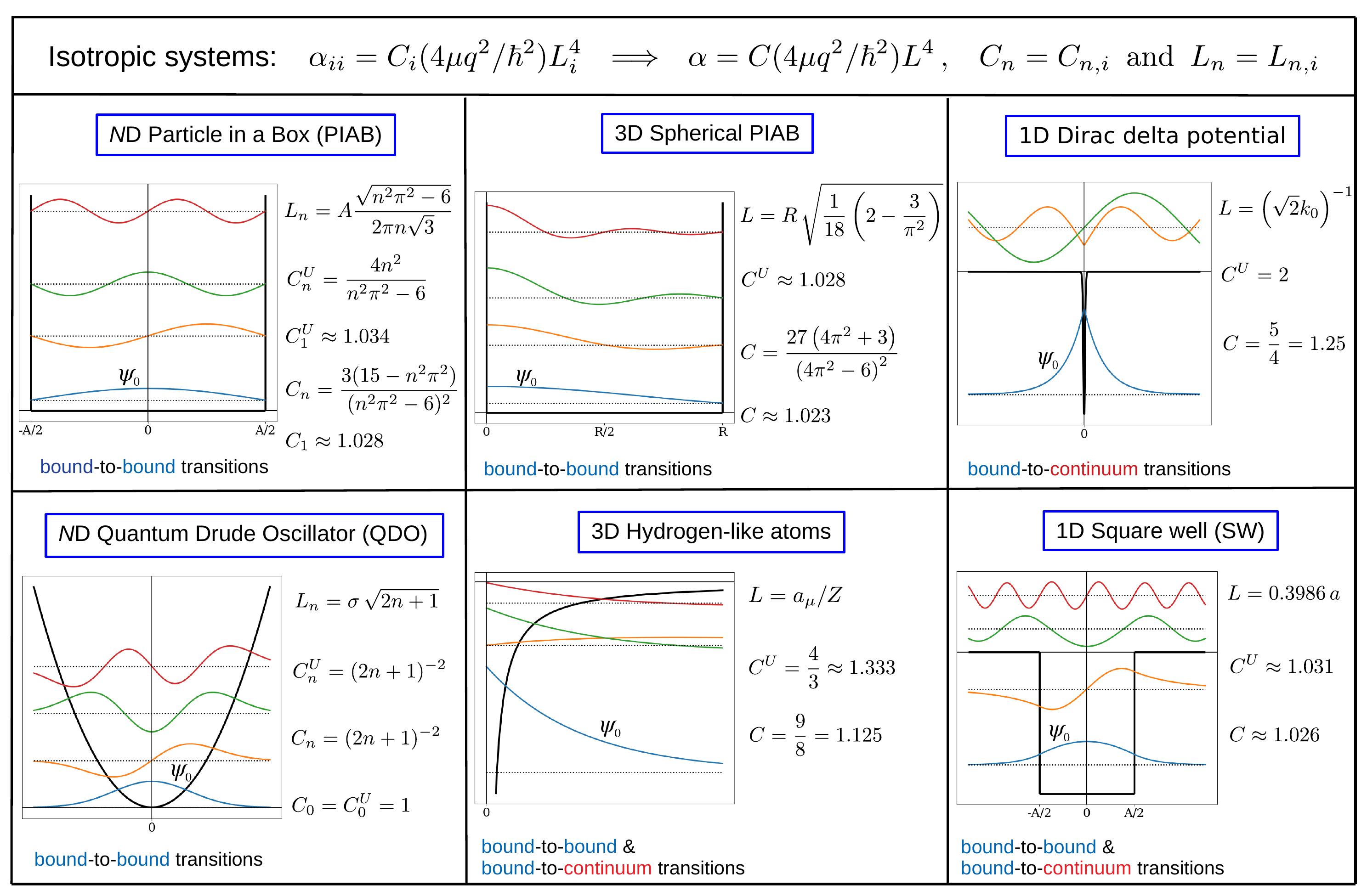}
\caption{The exact and Uns\"old ($U$) polarizability of six different quantum-mechanical models is represented by the general formula of Eq.~\eqref{eq.:alpha_L4}, where the characteristic length $L$ is calculated for each system according to the same unified definition given by Eq.~(\ref{eq.:L_definition}). The Cartesian indices ($ii$) are dropped since all the models are isotropic. For the $N$-dimensional ($N$D) PIAB and QDO, $n$ denotes the quantum number of excited states (with $n=1$ and $n=0$, respectively, for the ground state). For the other systems, only the ground state polarizabilty is evaluated. Furthermore, $A$, $k_0$, $R$, $r_0$ are the parameters of the potentials for the given systems~\cite{Supplementary}. 
Then, $\sigma^2$ is the variance of the harmonic oscillator, $Z e$ is the nuclear charge for the hydrogen-like atoms, and $a_\mu=(4\pi\epsilon_0)\hbar^2/\mu e^2$. The 1D square well is considered with the depth $V_0$ and the width $A$ related as $\displaystyle V_0=100\hbar^2/2\mu A^2$.} 
\label{img:Figmain}
\end{figure*}

To assess the scope of validity of Eq.~(\ref{eq.:alpha_L4}), we analysed several isotropic QM models: (i) particle in a box (PIAB) of an arbitrary dimension; (ii) particle confined in a sphere; (iii) 1D Dirac delta potential; (iv) square well (SW) in 1D; (v) quantum Drude oscillator (QDO) in an arbitrary dimension; (vi) 3D hydrogen-like atoms. These model systems are chosen since they allow one to obtain exact analytical solutions for their spectrum and polarizability containing a wide variety of features representing real molecules and materials. Figure~\ref{img:Figmain} summarizes the obtained polarizabilities for these models, whereas the detailed derivations are given in the Supplemental Material (SM)~\cite{Supplementary}. We show that each system obeys the formula given by Eq.~(\ref{eq.:alpha_L4}) with $L = L_i$ defined by Eq.~(\ref{eq.:L_definition}) for all models and constant $C = C_i$ being close to unity. This is remarkable since the chosen systems possess qualitatively different energy spectra: the PIAB and QDO have bound excited states only, while the 1D Dirac delta potential has solely excitations to the continuum; on the other hand, for the hydrogen-like atoms and SW, both bound and continuum states are present. Moreover, the polarizability of excited states of PIAB and QDO follow the same $L^4$ scaling law of Eq.~(\ref{eq.:alpha_L4}), as shown by a straightforward
generalization of our approach to an arbitrary QM state~\cite{Supplementary}. Thus, the polarizability of different models,
regardless of their spatial dimension, excitation state and spectra, can be expressed by Eq.~(\ref{eq.:alpha_L4}).

The difference between the model systems is reflected in both, the characteristic length and dimensionless constant entering Eq.~\eqref{eq.:alpha_L4}, but only $L$ contains the system parameters, while $C$ is independent of their actual values. As shown by Fig.~\ref{img:Figmain}, qualitatively similar systems have practically the same constant: $1.023 < C < 1.028$, for $N$D PIAB, 3D Spherical PIAB, and 1D SW (with $\displaystyle V_0=100\hbar^2/2\mu A^2$), as the cases of confined particles. For 1D SW, in the limit of vanishing potential depth, only one bound state remains resulting in $C=1.25$~\cite{Supplementary}, which is similar to 1D delta potential with just one bound state and $C=1.25$. For hydrogen-like atoms and the QDO, we obtain
$C = 1.125$ and $C=1$, respectively. 
As mentioned above, the Kirkwood approximation gives $C^{K}=1$ in all cases. Figure~\ref{img:Figmain} also shows $C^{U}$ obtained within the Uns\"old approximation, the upper bound of $C$. For the QDO, both approximations deliver the exact result, $C=1$. Thus, for atom-like systems in their ground state, $\Delta C$ in $C = 1 + \Delta C$ measures the strength of anharmonic contributions from quantum fluctuations versus the dominant harmonic part. The largest anharmonic contribution,
$\Delta C = 0.25$, is found for 1D delta potential and SW with vanishing potential depth~\cite{Supplementary}. 
Remarkably, for hydrogen-like atoms, $\Delta C = 0.125$ is exactly twice less than $\Delta C = 0.25$ obtained for the two systems with just one bound state. For the other three systems with confined particles, $\Delta C$ is vanishingly small, similar to the harmonic potential. For excited states, $C$ can strongly deviate from unity depending on corresponding quantum
numbers (see Fig.~\ref{img:Figmain} and the SM~\cite{Supplementary}), but the $L^4$ scaling remains valid. Altogether, this demonstrates that the general form of Eq.~\eqref{eq.:alpha_L4} is valid for all single-particle systems shown in Fig.~\ref{img:Figmain}. Furthermore, in the SM~\cite{Supplementary} we consider the nearly free electron model and show that the same scaling law holds for Bloch electrons.

\begin{figure}[t]
\includegraphics[width=9cm]{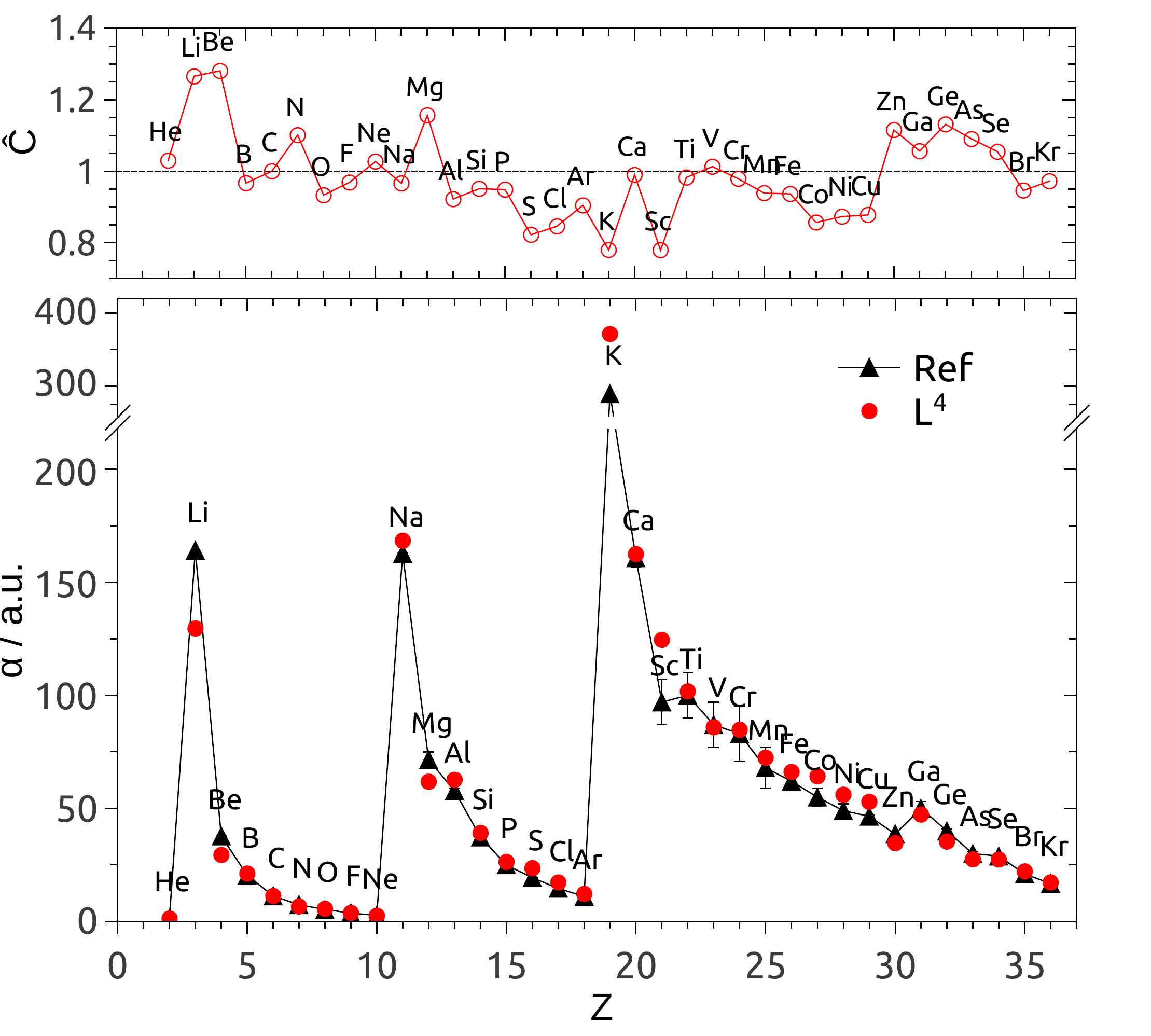}
\caption{Polarizabilities of multi-electron atoms calculated using ground-state DFT/PBE0 orbitals and the quantum-mechanical ($L^4$) scaling law. The lower panel demonstrates $\alpha$ calculated by Eq.~\eqref{Eq_L4_atomic} with ${\tilde C} = C_{k} = 1$. The upper panel shows ${\tilde C}$ obtained by comparison of Eq.~\eqref{Eq_L4_atomic} to the reference polarizabilities~\cite{Atomic_alpha_ref}.
Further details are given in the SM~\cite{Supplementary}.}
\label{img:Fig1}
\end{figure}

In the SM~\cite{Supplementary}, we discuss an extension of Eq.~\eqref{eq.:alpha_L4} to a general many-particle system. Here, we consider many-electron atoms by applying Eq.~\eqref{eq.:alpha_L4} to each electron shell. In such a case, the (isotropic) atomic polarizability reads
\begin{equation}
    \alpha = \tfrac{4 m_e e^2}{\hbar^2} \textstyle\sum\limits_k^{\text{occ}} \frac{C_{k}}{\eta_{k}} \frac{L_{k}^4}{N_k} \approx {\tilde C} \left(\frac{4 m_e e^2}{\hbar^2}\right) \textstyle\sum\limits_k^{\text{occ}} \frac{L_{k}^4}{\eta_{k} N_k}\ ,
\label{Eq_L4_atomic}
\end{equation}
where the sum runs over occupied orbitals with degenerate states treated together~\cite{Supplementary}, $L_{k}$ is obtained by Eq.~\eqref{eq.:L_definition} for  the $k$th orbital and $N_k$ is its occupation number stemming from the many-electron version of the TRK sum-rule. Then, $\eta_k$ are orbital-dependent factors required for all atoms starting from Li ($\eta_k^{_{\rm He}} = 1$), empirically found by us to be $\eta_k=n_k^{_{\ell}}N_k^{_{[1+(-1)^\ell]/2}}$, where $\ell$ and $n_k$ are, respectively, the orbital and principal quantum numbers of the $k$th orbital~\cite{Supplementary}. 
Based on our results for single-particle models, we assume all $C_{k}$ to be close to each other, which allows us to make the approximation given by the r.h.s.~of Eq.~\eqref{Eq_L4_atomic}.
As shown in Fig.~\ref{img:Fig1}, the coarse-grained constant ${\tilde C}$ is close to unity for different atoms. Hence, the response of each electronic orbital in an atom is well approximated by Eq.~\eqref{Eq_L4_atomic} with $C_k=1$. This means that we approximate each orbital in a many-electron atom by an effective quantum harmonic oscillator, where the screening of nuclear charge caused by the presence of occupied orbitals is taken into account via $\eta_k$.
In particular, ${\tilde C} \approx 1$ for noble gases from He to Kr, for which the QDO model is known to work well~\cite{Jones2013,Fedorov2018}.
Unlike single-particle systems, for many-electron atoms ${\tilde C}$ can be lower than unity, and we attribute this to correlation effects between electronic shells that should be explicitly included for a more accurate treatment. Equation~\eqref{Eq_L4_atomic} significantly improves over the many-electron version of the Kirkwood approximation derived by Buckingham~\cite{Buckingham1937}, which corresponds to setting $C_k/\eta_k=1$ in Eq.~\eqref{Eq_L4_atomic} and treating all
orbitals equally. Such approximation yields an overestimation
up to a factor of 4 for the polarizabilities shown in
Fig.~\ref{img:Fig1}~\cite{Supplementary}. 

\begin{figure}[t]
\includegraphics[width=8.5cm]{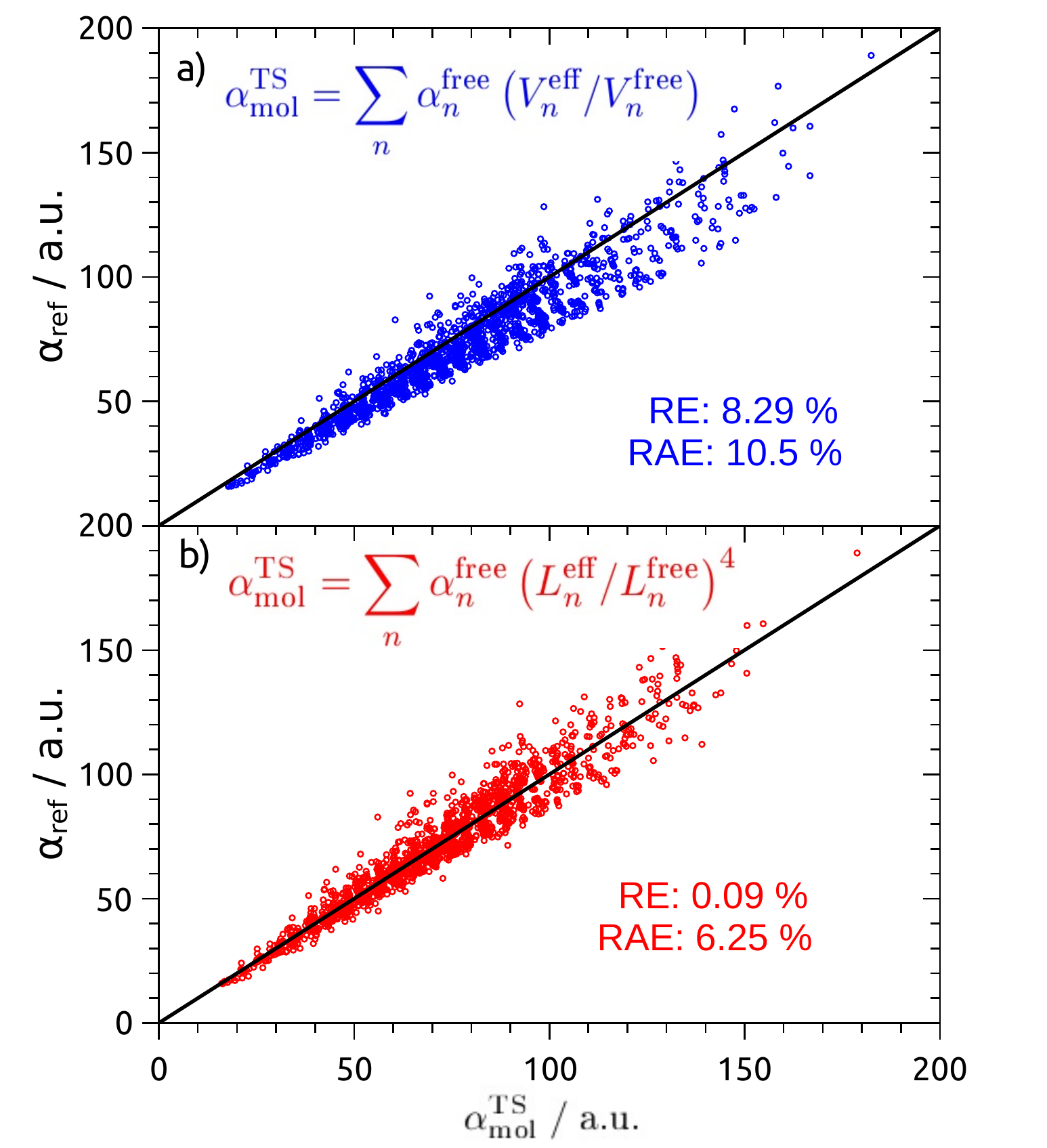}
\caption{Comparing two implementations of the Tkatchenko-Scheffler (TS) method~\cite{Tkatchenko2009}:
(a) the conventional TS method based on Eq.~\eqref{eqHirshfeld} and (b) the modified TS method based on Eq.~\eqref{eqHirshfeld_L}. The corresponding molecular polarizabilities $\alpha_{\rm mol}^{\rm{TS}}$ are shown versus the reference values $\alpha_{\rm{ref}}$ calculated by using density-functional theory (DFT) with PBE0 functional~\cite{Comment_2}.}
\label{img:Fig2}
\end{figure}

Let us now apply Eq.~(\ref{eq.:alpha_L4}) to compute polarizabilities of 1641 small organic molecules from the TABS dataset~\cite{Blair2014_1} by employing the Tkatchenko-Scheffler (TS) method~\cite{Tkatchenko2009}, which is widely used for vdW-inclusive density-functional calculations. Due to the commonly assumed direct proportionality between the atomic volume and polarizability, within the TS method molecular polarizabilities are approximated by a sum of effective atomic polarizabilities expressed in terms of the polarizabilities of free atoms as
\begin{equation}
\label{eqHirshfeld}
\alpha_{\rm mol}^{\rm TS} = \textstyle\sum\limits_n \alpha_n^{\rm eff}
= \sum\limits_n \alpha_n^{\rm free} \left({V_{n}^{\rm{eff}}}/V_{n}^{\rm{free}}\right)\ ,
\end{equation}
where the sum runs over all atoms in the molecule. The weights $\left({V_{n}^{\rm{eff}}}/V_{n}^{\rm{free}}\right)$ measuring
the volume ratio for atom in a molecule to the free atom in vacuum are obtained by the Hirshfeld partitioning of the electron density~\cite{Hirshfeld1977}. Based on the relation of Eq.~\eqref{eq.:alpha_L4}, we modify Eq.~\eqref{eqHirshfeld} to
\begin{equation}
\label{eqHirshfeld_L}
\alpha_{\rm mol}^{\rm TS} = 
\textstyle\sum\limits_n \alpha_n^{\rm eff}
= \sum\limits_n \alpha_n^{\rm free} 
({L_{n}^{\rm{eff}}}/L_{n}^{\rm{free}})^4\ ,
\end{equation}
which allows us to keep the simplicity of the TS method but make it consistent with the $L^4$ scaling law. Figure~\ref{img:Fig2} shows that by using Eq.~\eqref{eqHirshfeld_L} instead of Eq.~\eqref{eqHirshfeld} the average signed relative error $\langle {\rm RE} \rangle$ drops from 8.29\% to 0.09\%. The practically vanishing systematic deviation and the decrease of the average absolute relative error $\langle {\rm RAE} \rangle$ from 10.5\% to 6.25\% confirm  the applicability of the employed scaling law. The remaining deviations from reference DFT results can be attributed to the anisotropy of molecular polarizability, which necessitates an explicit coupling between atomic polarizabilities~\cite{Tkatchenko2012}. A detailed analysis performed in the SM~\cite{Supplementary} shows that among other possible scaling laws Eq.~\eqref{eqHirshfeld_L} provides the best accuracy for the TS method, which serves as an additional argument~\cite{Comment_3} for the general validity of Eq.~(\ref{eq.:alpha_L4}).

In summary, we have established a general formula for the dipole
polarizability, $\alpha = C (4 \mu q^2/\hbar^2)L^4$, valid for
QM systems of varying spatial dimension, symmetry, excitation state, and number of particles.~The~universality of the $L^4$ scaling for $\alpha$ is connected to the unified QM metric $L$ measuring fluctuations of the particle position in terms of the system parameters. By contrast, the dimensionless coefficient $C$ reflects just the qualitative properties of the eigenvalue spectrum of each system. The geometric scaling of the polarizability for a system in its ground state is solely determined by the ground-state wavefunction, whereas the effect of excited states is encoded in $C$ only. 
Another interesting finding is that the polarizability expression in Eq.~(\ref{eq.:alpha_L4}) is directly proportional to the particle mass, which is opposite to the classical picture where the polarizability vanishes for infinite particle mass. 
Our formula can be used to improve DFT-based methods for vdW
interactions~\cite{Hermann2017,Tkatchenko2009}, parametrize polarizable force fields~\cite{Wang2001,Sommerfeld2005,Jones2013}, or efficiently calculate dynamic spectroscopic observables based on the polarizability (\textit{i.e.}, Raman and sum-frequency generation)~\cite{Wang2006,Dakovski2007,Seufert2001,Empedocles1997,Kulakci2008}. These applications rely on efficient and accurate evaluation of polarizability from ground-state electron density. 

The authors acknowledge financial support from the Luxembourg National Research Fund: FNR CORE projects \lq\lq QUANTION(C16/MS/11360857, GrNum:11360857)\rq\rq\ and
\lq\lq PINTA(C17/MS/11686718)\rq\rq , AFR PhD Grant \lq\lq POMO(AFR PhD/19/MS, GrNum:13590856)\rq\rq , and \lq\lq DRIVEN\rq\rq\ (PRIDE17/12252781) under the PRIDE program.
Furthermore, the financial support from European Research Council, ERC Consolidator Grant \lq\lq BeStMo(GA n725291)\rq\rq\ and INTER-FWO project \lq\lq MONODISP\rq\rq , is also gratefully acknowledged.


\begin{thebibliography}{99}
\bibitem{IMVT1} The integral mean value theorem~\cite{IMVT2}: If $f(x)$ is a continuous real function on $x\in\left[a,b\right]$ and $g(x)$ is Riemann integrable on $[a,b]$ such that either $g\left(x\right)\geq0$ or $g\left(x\right)\leq0$, then there exists a real number $\xi\in\left[a,b\right]$ such that $\intop_{a}^{b}f\left(x\right)g\left(x\right)\text{d}x = f\left(\xi\right)\intop_{a}^{b}g\left(x\right)\text{d}x$.
\end{thebibliography}
\end{document}